\documentclass[a4paper]{article}  

\usepackage{color}
\usepackage{amsmath}
\usepackage{bbold}

\usepackage{mathtools}
\usepackage{amssymb}
\usepackage{graphicx}
\usepackage{gensymb}
\usepackage{pdfpages}
\usepackage{enumitem}
\usepackage{pifont}
\usepackage{MnSymbol}
\usepackage{hyperref}
\usepackage{epsfig}

\definecolor{drkgreen}{rgb}{0.0, 0.5, 0.0}

\DeclarePairedDelimiter\bra{\langle}{\rvert}
\DeclarePairedDelimiter\ket{\lvert}{\rangle}
\DeclarePairedDelimiter\avg{\langle}{\rangle}
\DeclarePairedDelimiterX\braket[2]{\langle}{\rangle}{#1 \delimsize\vert #2}

\title{Residual entropy and critical behavior of two interacting boson species in a double well}

\author{Fabio Lingua $^{1}$, Andrea Richaud $^{2,*}$ and Vittorio Penna $^{2}$\\
{\footnotesize $^{1}$ \quad Department of Physics, Clark University, Worcester, Massachusetts 01610, USA;}\\
{\footnotesize $^{2}$ \quad  Department of Applied Science and Technology and u.d.r. CNISM, Politecnico di Torino, I-10129 Torino, Italy}\\
 {\footnotesize $^{*}$ \quad Correspondence:   {\tt \footnotesize andrea.richaud@polito.it}
}}

\begin{document}
\maketitle

\abstract{Motivated by the importance of entanglement and correlation indicators in the analysis of quantum systems, we study the equilibrium and the bipartite residual entropy in a two-species Bose Hubbard dimer when the spatial phase separation of the two species takes place. We consider both the zero and non-zero-temperature regime. We present different kinds of residual entropies (each one associated to a different way of partitioning the system), and we show that they strictly depend on the specific quantum phase characterizing the two species (supermixed, mixed or demixed) even at finite temperature. To provide a deeper physical insight into the zero-temperature scenario, we apply the fully-analytical variational approach based on su(2) coherent states and provide a considerably good approximation of the entanglement entropy. Finally, we show that the effectiveness of bipartite residual entropy as a critical indicator at non-zero temperature is unchanged
when considering a restricted combination of energy eigenstates.}

\section{Introduction}
Systems formed by gases of ultracold bosons trapped
in homogenous arrays of potential wells (optical lattices) \cite{AP315} have attracted, in the last two decades, an enormous attention due to the rich variety of phenomena they feature at zero temperature \cite{LP19,RMP80}. The physical properties of such {\it quantum fluids} have been shown to be mainly determined by the competition of boson-boson repulsive interactions with the tunneling effect between adjacent wells, causing the boson mobility through the lattice. Among many effects observed in such systems, one of the most significant is the famous superfluid-insulator transition in which, for a boson-boson interaction strong enough, the boson mobility is quantum-mechanically inhibited when the boson density takes integer values \cite{PRB40}.

In this framework,
introducing in the lattice a second bosonic species interacting with the primary one has allowed the realization \cite{mix1,mix2} of the so-called {\it binary} quantum fluids whose complex phenomenology has revealed unexpected effects and behaviors. These are, for example, the formation of new types of insulating phases and superfluidity \cite{PRL90,PRA72}, quantum emulsions exhibiting a glassy character \cite{PRL98,PRL100}, the deformation of the insulating (Mott) domains accompanying the formation of polaron excitations \cite{PRA82,PRA89}, the presence of interspecies entanglement \cite{NJP16}, and the spatial separation of the two species (demixing effect) \cite{PRA88,PRA93}.

The simplest possible lattice system in which the interplay of two species can be studied is represented by the two-species dimer (TSD), namely, a mixture trapped in a lattice with two wells (dimer). This system, sufficiently simple to allow the use of standard analytic approaches, is however complex enough to exhibit the space-localization effects distinguishing two-species mixtures in larger lattices at zero temperature. The TSD has allowed both a thorough study \cite{JPB49} of such behaviors when the interspecies (repulsive) interaction $W>0$ is varied and the analytic derivation of the critical value of $W$ for which the mixed species, equally distributed in the two wells, localize in two separated wells. Such an effect, called delocalization-localization (DL) transition, also takes place in the attractive case ($W<0$) but in the final state both species occupy the same well for $|W|$ strong enough.
The DL transition, characterizing the ground state of the TSD, and the spectral collapse related to this phenomenon have been studied numerically in Ref. \cite{JPB49} and by means of the continuous-variable approach in Ref.  \cite{PRE95}. The critical behavior of the TSD has been confirmed by resorting to quantum-correlation indicators such as the Fisher information, the coherence visibility and the entanglement entropy (EE). The latter, in particular, has proved particularly sensitive in detecting the macroscopic changes in the ground-state structure both for repulsive and for attractive interspecies interaction.

This motivates our interest for the entanglement entropy and, more in general, for the {\it bipartite residual entropy} at non-zero temperatures in the TSD. In this paper, we perform a systematic study of this correlation property effecting numerical simulations which include non-zero temperatures. We begin with studying the zero-temperature regime. To check the robustness of the EE, we determine its dependence from the interspecies interaction both by considering the exact ground state (calculated numerically) and by representing the ground state in terms of atomic coherent states (CS). The CS picture is interesting since, in addition to allow for fully-analytic calculations, its semiclassical character approximates the system ground state in a form closer to the preparation of the system in real experiments.
As is well known, the EE describes the entanglement property of a physical system through the Von Neumann entropy of a suitably defined sub-system. In the sequel, we calculate the EE by partitioning the system in sub-systems such as \textit{i)} the left-well and the right-well bosons, \textit{ii)} bosons with zero and non-zero momentum,  and \textit{iii)} the species-A and species-B bosons.

As noted above, a number of new quantum phases has been predicted in the last fifteen years whose distinctive feature is to manifest at zero temperature. On the other hand, after the realization of optical lattices trapping ultracold atoms, it has become more and more evident that reducing (and measuring) the temperature on the nanoscale is an outstanding problem \cite{RPP74}. For this reason, the detection of zero-temperature phase transitions such as the space separation in bosonic mixtures (or its simpler dimer version, the DL transition) must more realistically rely on indicators which are reminescent of the critical behavior of the system even when temperature is non-zero \cite{SR5}.

In this perspective, achieving a good control of the correlation properties for a system undergoing the DL transition at non-zero temperature certainly represents a useful tool for its observation in future experiments. For this reason we have thoroughly explored the residual-entropy behavior at non-zero temperature and have tried
to understand the effect of thermal fluctuations in regimes where they compete with
quantum fluctuations. The bipartite residual entropy has been calculated numerically by exploiting the knowledge of the TSD exact spectrum. As in the zero-temperature case, we consider the reduced thermal density matrix for three different partition schemes of the system based on separating space modes, momentum modes and atomic species.
Finally, to further test the bipartite residual entropy as a critical indicator, we have compared the exact bipartite residual entropy with that calculated using a restricted range of energy levels around the expected average energy.

The paper is organized as follows. In Section \ref{sec:the_model}, we introduce the TSD model and review the DL transition discussing the change of structure it induces in the ground state and the spectral collapse, a significant property that marks the transition. Section \ref{sec:S_eq_S_R} is devoted to defining the equilibrium and the bipartite residual entropy and the relation thereof with the EE. Section \ref{sec:SR_T} contains the results of our numerical calculations of bipartite residual entropy within the previously discussed partition schemes at zero and non-zero temperature. In Section \ref{sec:Coherent_States} and Section \ref{sec:Restricted} we compute the bipartite residual entropy in the coherent-state and in the restricted-basis approach, respectively. Section \ref{sec:Conclusions} is devoted to concluding remarks.


\section{The model and the ground-state properties}
\label{sec:the_model}
An effective description of ultracold bosons trapped in homogenous arrays of potential wells is provided by the Bose-Hubbard (BH) model \cite{PRB40} in which local boson operators $A_i$ and $A^+_i$ represent the microscopic annihilation and creation processes, respectively, at the $i$th well.
The experimental realization of this model is currently achieved by means of well-known optical-trapping techniques \cite{RMP80,RMP78}. These, by combining counter-propagating laser beams, cause the formation of (optical) lattices the sites of which correspond to effective local potentials attracting bosons. In the simplest possible case of a two-site lattice (a double potential well), the BH Hamiltonian is given by

$$
H_a =
\frac{U_a}{2}\Bigl [ A_{L}^{\dagger} A_{L}^{\dagger} A_{L} A_{L}
+
A_{R}^+ A_{R}^+ A_{R} A_{R} \Bigr ]
-J_a \big( A_{L}^+ A_{R}+ A_{R}^+ A_{L} \big ),
$$
where $L$ and $R$ refers to the left and right well, respectively, $U_a$ is the boson-boson interaction and $J_a$ is the hopping amplitude controlling interwell boson exchange. The boson operators $A_L$, $A_L^+$, $A_R$, and $A_R^+$ satisfy the standard commutator $[A_\sigma, A_\sigma^+]$ $=1$ with $\sigma= L,R$.
If, in addition to species A a second species B is introduced, the spatial modes become four, $A_L$, $A_R$, and $B_L$, $B_R$, for the species $A$ and $B$, respectively. The resulting mixture is thus described by the two-species dimer Hamiltonian \cite{JPB49}

\begin{equation}
\label{model}
{H} ={H}_{a}+ {H}_{b} + W \big ( N_L M_L +N_R M_R \big )
\end{equation}
in which, apart from the single-species BH Hamiltonians ${H}_{a}$ and ${H}_{b}$, the significant term is that depending on interspecies interaction $W$. This couples the two species through the boson local populations described by the number operators
$N_\sigma = A_\sigma^+A_\sigma$ and $M_\sigma = B_\sigma^+ B_\sigma$ with $\sigma= L,R$.

When the interspecies interaction $W$ becomes sufficiently strong, the two interacting species trapped in a double-well potential feature macroscopic localization effects. In particular, a repulsive interaction tends to spatially separate the species into different wells while an attractive interaction tends to confines both species in the same well. This represents the DL transition. In the first case this is characterized by an almost complete localization of the two species in different wells, and thus by a demixing effect, whereas, in the second case, the attractive interaction leads to a ``supermixed" state with a localization of both species in a single well.

Such effects are confirmed by the numerical calculation of the ground state for different values of $W$. To see this we note that the energy eigenstates can be suitably represented in the basis of space-mode Fock states

\begin{equation}
| n_L , m_L, n_R, m_R \rangle  := | i , j \rangle_L |N-i, M-j \rangle_R, \qquad i\in [0,N], \,\, j\in [0,M],
\label{FBS}
\end{equation}
where labels $n_\sigma$ and $m_\sigma$, describing the local boson populations, are the eigenvalues of number operators $N_\sigma$ and $M_\sigma$, respectively.
The parametrization $n_L = i$, $m_L = j$, $n_R = N-i$ and $m_R = M-j$ has been assumed to include the property that both operator $N = N_L + N_R$ and operator $M = M_L +M_R$ (representing the total boson numbers of the two species) commute with Hamiltonian $H$ and thus are conserved quantities. The factorized form of (\ref{FBS}) aims to better distinguish left-well from right-well populations. A generic quantum state is then represented as

\begin{equation}
| \Psi \rangle = \sum^N_{i=0} \sum^M_{j=0} w_{i,j}\;
\ket{i,j}_L \ket{N-i,M-j}_R
\label{E0}
\end{equation}
Determining the energy eigenstates thus amounts to calculating coefficients $w_{i,j}$ for which the eigenvalue equation $H |E \rangle= E |E \rangle$ is fulfilled.
For values of $W$ small enough, the ground state $| E_0 \rangle$ is approximated  in terms of su(2) coherent states \cite{JPA41}

$$
| E_0 \rangle \simeq \frac{1}{2^{(N+M)/2}\sqrt{ N! M!}} \left ( A^+_L + A_R^+ \right )^N \left ( B^+_L + B_R^+ \right)^M
| 0, 0\rangle_L |0, 0 \rangle_R
$$
whose dominating components $| i, j\rangle$ can be shown to feature $i \simeq N/2$, $j \simeq M/2$, namely, boson populations equally distributed in the two wells (delocalized ground state). For large values of $|W|$, $| E_0 \rangle $ can be approximated by

\begin{equation}
| E_0 \rangle \simeq \frac{1}{\sqrt 2}\Bigl ( | N , 0 \rangle_L | 0, M \rangle_R +
| 0 , M \rangle_L |N, 0 \rangle_R \Bigr ), \quad
| E_0 \rangle \simeq \frac{1}{\sqrt 2}\Bigl ( | N , M \rangle_L | 0, 0 \rangle_R +
| 0 , 0 \rangle_L |N, M \rangle_R \Bigr ),
\label{eq:cats}
\end{equation}
in the repulsive and attractive case, respectively, well illustrating the space-localized distributions emerging from the delocalization-localization transition \cite{JPB49} and leading to Schr\"odinger cats with strongly localized component states.

\begin{figure}[h]
\centering
\includegraphics[width=\textwidth ]{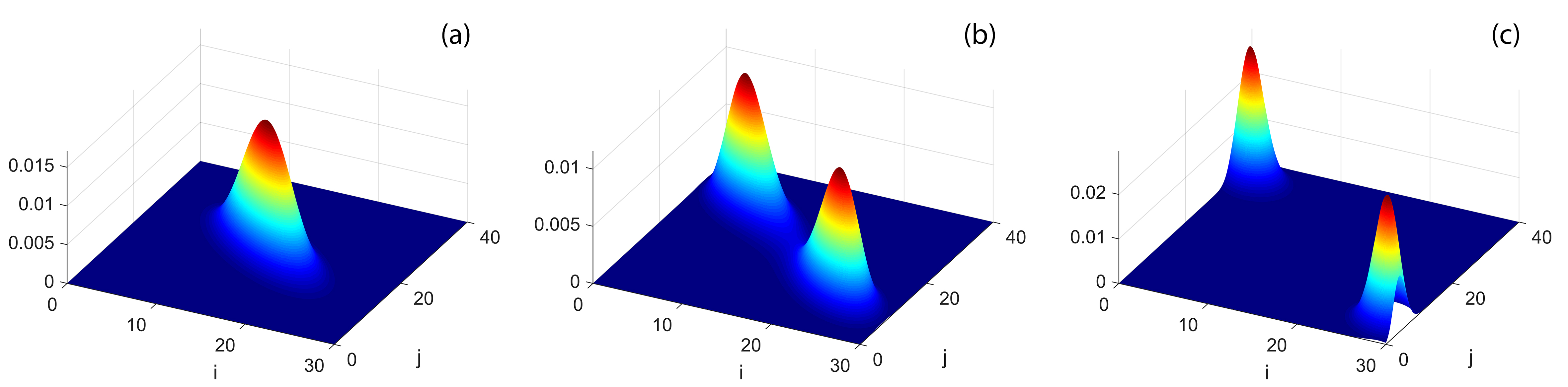}
\caption{(Color online)
Ground-state probabilities $|w_{i,j}|^2$ vs $i$ (left occupation number of species A) and $j$ (left occupation number of species B) associated to space-mode Fock states
$\ket{i,j}_L\ket{N-i,M-j}_R$ of equation (\ref{E0}) for boson numbers $N=30$, $M=40$ and $U=0.1$. Panel (\textbf{a}) features localized populations for $W=0.15$, (\textbf{b}) partially
localized populations for $W=0.168$, and (\textbf{c}) fully separated populations for $W=0.2$. Energies in units of $J_a=J_b= J$.}
\label{fig1}
\end{figure}

Figure \ref{fig1}, obtained by numerically calculating the ground state in the repulsive case for different $W$, supplies us with an exact description of the DL transition and of the macroscopic changes in the ground-state structure. A similar behavior characterize the DL transition in attractive case, but the two emerging peaks finally localize around $i=j=0$ an $i=30$,
$j=40$.

The critical behavior of the DL transition has been studied analytically by resorting to the semiquantum approach where boson number operators are approximated in terms of continuous variables \cite{PRE95}. This method has provided the critical value of $W$ at which the transition takes place in the case of twin species ($J_a =J_b=J$, $U_a=U_b=U$). In this approach the Fock states essentially become wave functions depending on the new continuous variables while, for energies low enough, the energy-eigenvalue equation takes the form of the Schr\"odinger problem for a multidimensional harmonic-oscillator Hamiltonian. The extremal points of the corresponding potential allow one to determine the ground-state configuration, and, in particular, to find the formula

$$
W = 2J/N + U
$$
defining, for large boson numbers ($N=M \gg 1$), the transition critical point in the parameter space. Interestingly, when $W$ approaches this critical value, the energy spectrum has been shown to undergo a collapse in which the inter-level separation tends to zero. This spectral collapse can be seen as the hallmark of the dynamical transition which features the macroscopic change in the structure both of the ground state (see the previous discussion) and, more in general, of the low-energy excited states described in Ref. \cite{PRE95}. The generalized version of the previous formula for a mixture in a $L$-well ring lattice has been derived in Ref. \cite{PRA96}.


\section{Equilibrium entropy and bipartite residual entropy}
\label{sec:S_eq_S_R}
The third law of thermodynamics states that a perfect crystal at temperature $T=0$ exhibits entropy $S=0$. This entropy is defined as the \emph{Equilibrium Entropy} $S_{eq}$. However, several physical systems ranging from, e.g., water ice \cite{SR_ice_dimarzio,SR_ice_Lieb}, carbon monoxide \cite{SR_CO}, highly pressurized liquid-helium \cite{SR_He}, glass systems \cite{SR_glass,SR_review_glass}, proteins \cite{SR_proteins}, and even black-holes \cite{SR_BH_cov,SR_BH}, seems to manifest a residual content of information (corresponding to a residual entropy) for $T\rightarrow 0$.
The presence of such \emph{residual entropy} $S_R$ has been generally associated with residual degrees of freedom at $T=0$ such as, among others, ground-state degeneracy, residual structural disorder, geometrical frustration and entanglement. These physical phenomena act as sources of uncertainty preventing the possibility to acquire knowledge on the exact state of the system, thus resulting as possible sources of information (i.e. a finite, residual value of the entropy). 

For quantum systems, the residual entropy can be related to the presence of entanglement in the ground-state through the entanglement entropy\footnote{As an example, consider a two-spin system such that its ground state is
$|\Psi_0\rangle= (|\uparrow\,\downarrow\rangle + |\downarrow\,\uparrow\rangle)/\sqrt{2}$ ( $\hat{H}|\Psi_0\rangle = E_0|\Psi_0\rangle$ ). If, on one hand, $|\Psi_0\rangle$ is a pure state with no information at all if measured in the energy basis, on the other hand, the outcome of a measurement is not certain anymore if the spins of the two particles are measured. In that case a $50\%$ chance to measure either $|\uparrow\,\downarrow\rangle$ or $|\downarrow\,\uparrow\rangle$ there exists, hence suggesting a residual information content of one bit at $T=0$.
For this reason in quantum systems, the entanglement measure in the ground state by means of the entanglement entropy is associated to a bipartite residual entropy. }. 
Entanglement entropy is a measure of the ``amount of entanglement'' in the system.  A standard and accepted way to quantify entanglement is through the Von-Neumann entropy. What is measured by the bipartite entanglement entropy is the mutual information shared between two partitions of the physical system (e. g. Alice and Bob). Given $\hat{\rho}$ the density matrix of the system, and defining two partitions $A$, $B$ of the Hilbert space $\mathcal{H}$ such that $\mathcal{H}=\mathcal{H}_A\bigotimes\mathcal{H}_B$, the bipartite Von-Neumann entropy is defined as \cite{NielsenCuang}

\begin{equation}
S(\hat{\rho}_A)=-\mathrm{Tr}_A(\,\,\hat{\rho}_A\,\, \log_2 \,\,\hat{\rho}_A),
\label{VonNeuS}
\end{equation}
where $\hat{\rho}_A=\mathrm{Tr}_B(\,\,\hat{\rho})$ ( $\hat{\rho}_B=\mathrm{Tr}_A(\,\,\hat{\rho})$) is the reduced density matrix of partition $A$ ($B$) obtained tracing out the degrees of freedom of $B$ ($A$). From now on we will refer to this indicator as \textit{bipartite residual entropy}. Notice that for the same system, in principle, there exists infinitely many possible ways to partitions the Hilbert space $\mathcal{H}$ in two parts. This leads to the consideration that, since the choice of the partition is arbitrary, the measure of entanglement, i.e. the bipartite residual entropy, cannot have a global character by definition. We shall see how this is indeed the case in Section \ref{sec:SR_T} (and, more specifically, in Subsection \ref{sec:SR_0}) when we will compute the bipartite residual entropy for the TSD for different choices of the partition $A$-$B$.

\subsection{Equilibrium Entropy in the TSD}
According to quantum statistical mechanics \cite{FeynmStatM,Landau}, the expression of the equilibrium entropy $S_{eq}(T)$ can be derived from the expression of the density operator as

\begin{equation}
S_{eq}(T)=-\mathrm{Tr}(\,\,\hat{\rho}\,\,\log_2\,\,\hat{\rho}) \label{Seq1}
\end{equation}
where the (canonical) density operator \cite{AmicoRMP} at finite temperature is defined as

\begin{equation}
\hat{\rho}=\frac{1}{\mathcal{Z}}\sum_n e^{-\beta E_n}  |\Psi_n\rangle\langle\Psi_n|\,\,,\label{Stat_densOp2}
\end{equation}
with $E_n$ representing the energy eigenvalue associated to the energy eigenstate $|\Psi_n\rangle$. Combining Equations (\ref{Seq1}) and (\ref{Stat_densOp2}) one finds the explicit expression for the equilibrium entropy

\begin{equation}
S_{eq}(T)=\sum_n -\rho_n \log_2 \rho_n\,\,, \label{Seq}
\end{equation}
with $\rho_n= e^{-\beta E_n}/\mathcal{Z}$.

\begin{figure}[h]
    \centering
    \includegraphics[width=0.9\columnwidth]{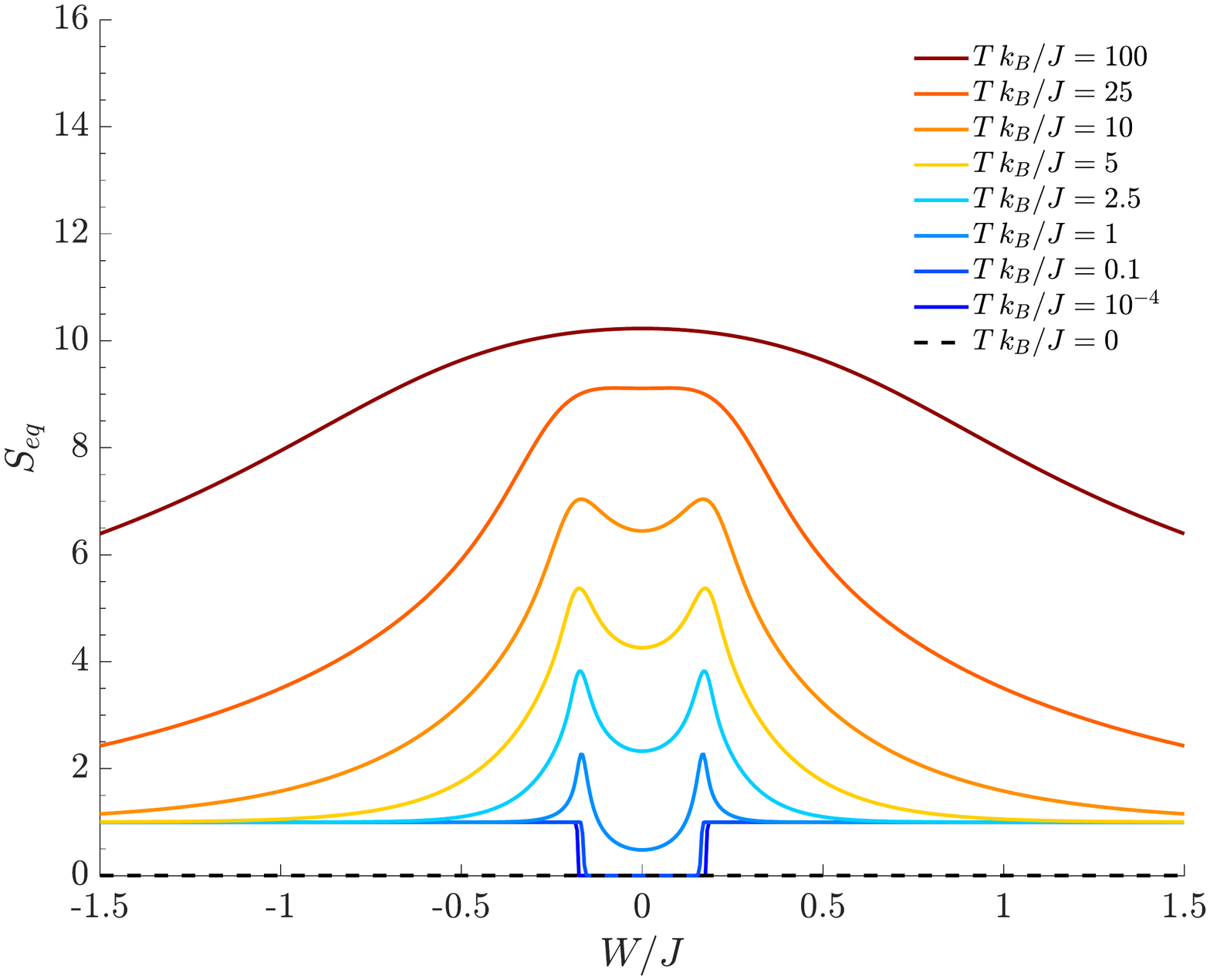}
    \caption{Equilibrium entropy for different choices of the temperature. $N=30$, $M=40$, $J=1$, $U=0.1$, $k_B=1$. }
   \label{fig:Equilibrium_entropy}
\end{figure}

Since the ground-state of Hamiltonian (\ref{model}) cannot be degenerate \cite{Landau, Wei2014}, expression (\ref{Seq}) is a good definition of equilibrium entropy for the TSD as, for $T=0$, it exactly satisfies $S_{eq}(0)=0$.
In Figure \ref{fig:Equilibrium_entropy}, we show the equilibrium entropy computed for the TSD as a function of the interspecies interaction $W/J$ and effective temperature $Tk_B/J$. At $T=0$ one clearly sees that $S_{eq}=0$ for all values of $W/J$ (black-dashed line). By sufficiently increasing the temperature, two peaks appear at the boundary of the central region where the phase transitions between the mixed and demixed phases occur ($|W|/J\approx 0.16$). Their presence is related to the energy-level spectral collapse: this, in fact, entails that a bigger and bigger number of energy levels significantly contributes to the thermal density matrix used to calculate this quantity.  Such peaks progressively vanish due to fluctuations when the temperature increases.
In Section \ref{sec:SR_T}, we will show that, in general, bipartite residual entropy exhibits similar features.

According to Figure \ref{fig:Equilibrium_entropy}, the equilibrium entropy tends to $S_{eq}= 1$ for $T\ne 0$
and if $|W|/J$ is large enough (plot tails). This reflects the fact that two dominating states (those corresponding to the lowest energies $E_1$ and $E_0$) provide contributions of about $\frac{1}{2} \log_2 2$ to the limiting value $S_{eq}= 1$. It is important to notice that, in both tails, the first excited level $E_1$ can be shown to tend to the ground state energy $E_0$ as a consequence of the spectral collapse characterizing the TSD. Accordingly, the smallest non-zero temperature that has been considered ($Tk_B/J=10^{-4}$) is large enough to populate in a nearly equal way both the ground state and the first excited level because their separation $E_1-E_0$ becomes smaller and smaller for large $|W|/J$.

Notice also that, in this regime, the splitting between $E_0$ and $E_1$ decreases exponentially (with the number of particles) to a point that may lie below the actual experimental limit (see Appendix \ref{appB} for details).
We also note that, at high temperatures, the equilibrium entropy tends to the value $S_{eq}\simeq 10.31 = \log_2 D$ where $D$ is the number of energy levels (i.e. the dimension of the Hilbert space), showing the fact that all the energy eigenstates are equiprobable with probability $1/D$.

\subsection{Bipartite residual entropy in the TSD}
\label{sec:Res_en_TSD}
In Reference \cite{JPB49} we showed that the TSD manifests non trivial entanglement properties (relevant to the boson distribution in the two wells) in the ground-state suggesting the presence of a bipartite residual entropy at $T=0$.
This residual information at $T=0$ is not grasped by (\ref{Seq}) as it exhibits $S_{eq}(0)=0$.

A consistent and different definition of the entropy is therefore required in order to be able to correctly describe the residual quantum information hidden in the ground-state structure. This can be naturally done by extending the definition of entanglement entropy (\ref{VonNeuS}) at finite $T$ in the way suggested by the expression for the equilibrium entropy (\ref{Seq}). We will call this quantity \emph{bipartite residual entropy at finite temperature} $S_R(T)$ in order to distinguish it from the equilibrium entropy $S_{eq}(T)$ of expression (\ref{Seq}).

The key difference between definitions (\ref{VonNeuS}) and (\ref{Seq1}) lies in the fact that, in the entanglement entropy, a reduced density operator $\hat{\rho}_A$ is used. Given a partition of the Hilbert space, the reduced density operator of $\mathcal{H}_A$ is obtained by tracing out the degrees of freedom of $\mathcal{H}_B$. The idea is then to compute the reduced density matrix of the thermal density operator $\hat{\rho}$ defined in (\ref{Stat_densOp2}), and then to use the new density operator for computing the bipartite residual entropy at finite $T$.
Although the partition of the Hilbert space is obviously independent from the choice of the basis in which the density operator is represented, to perform the calculation described above is convenient express the density operator (\ref{Stat_densOp2}) in an alternative suitable basis for the  partition $A$-$B$ one has chosen. From the practical point of view, a suitable choice of the basis can give easy access to a partition that in another basis would be really hard to handle computationally. An example of this is shown in Section \ref{sec:SR_T} when we consider the partition between the momentum modes.

Let's expand density operator (\ref{Stat_densOp2}) in a convenient basis $\{|\phi_i\rangle\}$ for the choice of the partition. To do so we expand the energy eigenstate

\begin{equation}
|\Psi_n\rangle=\sum_{i}\sum_{j} w_{i,j,n}\,\,|\phi_i\rangle_A \otimes |\phi_j\rangle_B\,\,,
\end{equation}
substitute it in expression (\ref{Stat_densOp2}), and obtain the new expression for the density operator \cite{KersonHuang}

\begin{equation}
\label{eq:thermal_superposition}
\hat{\rho}\equiv\hat{\rho}(T)=\sum_{i}\sum_{j}\sum_{i^\prime}\sum_{j^\prime} C_{i,j,i^\prime,j^\prime}(T)|\phi_i\rangle_A \otimes |\phi_j\rangle_B \,_B\langle\phi_{j^\prime}| \otimes  _A\langle\phi_{i^\prime}|\,\,,
\end{equation}
where

\begin{equation}
C_{i,j,i^\prime, j^\prime}(T)=\sum_n \frac{e^{-\beta E_n}}{\mathcal{Z}}\,\, w_{i,j,n} {w_{i^\prime j^\prime,n}^*}\,\,.
\label{eq:C_ij}
\end{equation}
Notice that, coefficients $C_{i,j,i^\prime, j^\prime}(T)$ contains both thermal and quantum information as they are obtained by thermal-averaging the quantum amplitudes $w_{i,j,n} w_{i^\prime, j^\prime,n}^*$  of each energy eigenstate $\ket{\Psi_n}$. By tracing over the degrees of freedom of $\mathcal{H}_B$ is possible to derive the expression of the reduced density operator $\hat{\rho}_A(T)$

\begin{equation}
\hat{\rho}_A(T)=\mathrm{Tr}_B(\hat{\rho}(T)),
\label{red_rho}
\end{equation}
The bipartite residual entropy at finite temperature $S_R(T)$ is then defined as

\begin{equation}
S_R(T)=-\mathrm{Tr}_A(\,\,\hat{\rho}_A(T)\,\,\log_2 \,\,\hat{\rho}_A(T))\,\,.
\label{SR}
\end{equation}
The details of this calculation, together with the results of the computation of (\ref{red_rho}) and (\ref{SR}) for different choices of the partition, are discussed in Section \ref{sec:SR_T}.

\section{Bipartite residual entropy at zero and finite temperature}
\label{sec:SR_T}
As already mentioned, ``bipartite entanglement" is well defined when the way to partition the system with respect to a certain physical property is specified. Investigating specific properties of a given system leads to consider specific kinds of entanglement. An effective and standard way to quantify the bipartite residual entropy is to compute the Von Neumann entropy according to the scheme discussed in the previous Section. Of course, once the partition is fixed, the computation of the Von Neumann entropy relevant to the reduced density matrix (bipartite residual entropy) is independent on the basis chosen to represent physical states, namely, $S(\rho)=S(U\rho U^\dagger)$ for any unitary transformation $U$ which enacts the change of basis. 

In the sequel, we consider three different kinds of bipartite residual entropy, each one associated to a different way of partitioning the system. First, we consider the quite natural partition in terms of left-well bosons and right-well bosons suggested by the representation of physical states in the space-mode Fock basis (\ref{FBS}). Then, by representing physical states in the momentum-mode Fock basis, we partition the system in terms of zero-momentum and non-zero-momentum bosons. Finally, we consider the partition of the system distinguishing species-A from species-B bosons, which is again suggested by definition (\ref{FBS}) where populations $n_L$, $n_R$ and $m_L$, $m_R$ refer to species A and B, respectively. In all three cases we present the results, obtained numerically, both for the zero-temperature scenario, when only the ground state $\ket{\psi_0}$ is involved and for the finite-temperature configuration, when the system is naturally described by means of a thermal density matrix. It is worth remarking that at $T=0$, the bipartite residual entropy reduces to entanglement entropy because classical correlations are suppressed.

\subsection{Bipartite residual entropy for a partition characterized by spatial modes}
\label{sec:Entanglement_LR}
\begin{figure}[h]
    \centering
    \includegraphics[width=0.9\columnwidth]{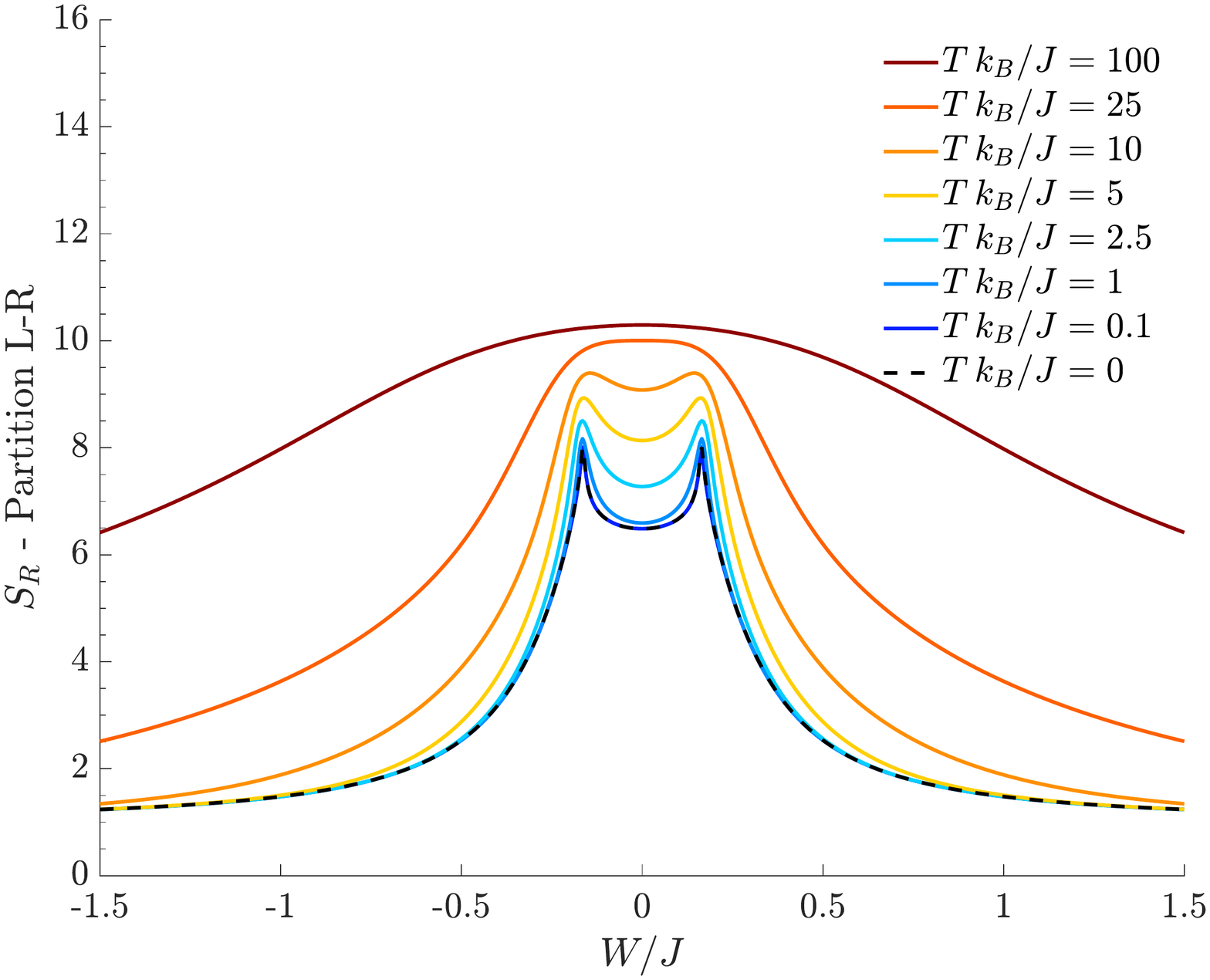}
    \caption{Bipartite residual entropy relevant to the L-R partition for  different choices of the temperature. $N=30$, $M=40$, $J=1$, $U=0.1$, $k_B=1$.}
\label{fig:S_R_LR}
\end{figure}

Let us start by computing the bipartite residual entropy $S_R$ by considering the partition of the TDS in terms of left-well bosons and the right-well bosons. Following Formula (\ref{E0}), a generic physical state $\ket{\psi}$ is written as

$$
  \ket{\psi}=\sum_{i=0}^{N}\sum_{j=0}^{M}\,\, w_{i,j}\,\, \ket{i,j}_L \,\, \ket{N-i,\,M-j}_R
$$
and entails the density matrix of the whole system

\begin{equation}
\hat{\rho}=\ket{\psi}\bra{\psi} = \sum_{i=0}^{N}\sum_{j=0}^{M}\sum_{i^\prime=0}^{N}\sum_{j^\prime =0}^{M}\,\, w_{i,j} w_{i^\prime,j^\prime}^* \,\, \ket{i,j}_L \,\, \ket{N-i,\,M-j}_R
\prescript{\null}{R}{\bra{N-i^\prime,\,M-j^\prime}} \,\, \prescript{\null}{L}{\bra{i^\prime,j^\prime}}.
\label{eq:Density_matrix_sites}
\end{equation}
The reduced density matrix relevant to the right-well bosons, obtained tracing out the degrees of freedom of the left-well bosons, is

$$
  \hat{\rho}_R = \sum_{k=0}^{N} \sum_{l=0}^{M} \, \prescript{\null}{L}{\bra{k,l}}\hat{\rho}\ket{k,l}_L= \sum_{i=0}^{N}\sum_{j=0}^{M} \,\, |w_{i,j}|^2  \,  \,\, \ket{N-i,\,M-j}_R   \,  \prescript{\null}{R}{\bra{N-i,\,M-j}}.
$$
In the presence of a non-zero temperature, the density matrix modifies taking into account the contributions of the whole energy spectrum. By following the scheme discussed in Subsection \ref{sec:Res_en_TSD}, as the $T\neq 0$ density matrix is diagonal, one can easily compute the bipartite residual entropy (\ref{SR}) finding

\begin{equation}
 S_R(\hat{\rho}_R)= -  \sum_{i=0}^{N}\sum_{j=0}^{M} \,\, |C_{i,j}(T)|^2 \log_2 |C_{i,j}(T)|^2,
 \label{eq:S_R_LR}
\end{equation}
with $C_{i,j}(T)$ given by Formula (\ref{eq:C_ij}). Figure \ref{fig:S_R_LR} shows how the bipartite residual entropy relevant to right-well bosons varies with respect to $W/J$, for different temperatures.
At $T=0$, the plot of $S_R$ (black dashed line), which represents the entanglement entropy, exhibits two sharp peaks where the DL transitions occur. In the region between the two peaks bosons are delocalized and the quantum fluids fully mixed, the left tail corresponds to supermixed states (states where both species are localized in a single well) and, eventually, the right tail is the region where the two species localize in different wells. Both tails feature a genuinely quantum behavior because the relevant ground states correspond to Schr\"odinger cats, in which the spatial separation gets more and more pronounced as $|W|/J$ increases (see Formula (\ref{eq:cats})). In fact, the entanglement entropy asymptotically tends to $1$, a value which is reminiscent of the double-edged structure of cat states (\ref{eq:cats}) because both their components contribute to Formula (\ref{eq:S_R_LR}) with $\frac{1}{2}\log_2 2$.
It is worth noticing that, at $T=0$, $S_R$ is always different from zero in that, even for noninteracting species ($W=0$), the presence of a non-zero $J$ couples the left and right modes of either species. As expected, one can show numerically that the height of the central minimum of $S_R$ decreases more and more (tending to zero) as the interwell hopping $J$ becomes smaller and smaller.

At temperature $T>0$, Figure \ref{fig:S_R_LR} shows that the bipartite residual entropy is still able to highlight the difference among mixed, demixed and supermixed phases thanks to the presence of the two peaks. In addition to the spectral collapse mechanism, the emergence of the two peaks in the proximity of the $W/J$ critical values is caused by the bigger and bigger number of Fock states significantly contributing to increase $S_R$ (the same interpretation holds for the two peaks characterizing $S_R$ with A-B partition (see Figure \ref{fig:S_R_AB})). This effect is suggested by the fact that the two Gaussian-like distributions (see Figure \ref{fig1}b) involve the maximum number of Fock states when they are about to merge and the transitions take place. 

The effect of a finite temperature is to smooth the DL phase transitions, an effect which can be clearly appreciated observing the decreasing sharpness of the peaks as T is increased. Interestingly, all the tails of the plotted curves tend to the limiting value $1$. For example, in the left tail ($W/J<0$), this means that, in $S_R$ one has $|C_{0,0}|^2=|C_{N,M}|^2\simeq 1/2$ while all the other $|C_{i,j}|^2$ are vanishingly small. The resulting $S_R=1$ follows from the fact that there exist two dominating macroscopic configurations $\ket{N,M}_L \ket{0,0}_R$ and $\ket{0,0}_L \ket{N,M}_R$ whose correlation is mainly due to quantum entanglement for $T\to 0$ but assumes a more and more classical character for higher temperatures. In the case of the right tail ($W/J>0$), the same effect is observed but the dominating components are $|C_{N,0}|^2$ and $|C_{0,M}|^2$. Note that, at fixed temperature, such configurations emerge provided that the interspecies interaction $|W|$ is strong enough to contrast the temperature-induced disorder. Of course, for a given value of $W/J$, one has larger residual entropies at higher temperatures in that increasing $T$ makes more and more energy eigenstates accessible in the thermal superposition ensuing from Formula (\ref{Stat_densOp2}).
We conclude by observing that, at high temperatures, in the central region around $W/J=0$, $S_R$ approaches the limiting value $\log_2 D\approx 10.31$, because $|C_{i,j}|^2 \simeq 1/D$ for all $(i,j)$  where $D=1271=(N+1)(M+1)$ (with $N=30$, $M=40$) is the dimension of the Hilbert space. This limiting situation reflects the fact that, at high temperatures, $S_R \to S_{eq}$ (see Figure \ref{fig:Equilibrium_entropy}).

\subsection{Bipartite residual entropy for a partition characterized by momentum modes}
Let us introduce the following momentum-mode operators obtained summing and subtracting usual site-mode operators

$$
   S_a = \frac{1}{\sqrt{2}}(A_L+A_R), \qquad  D_a = \frac{1}{\sqrt{2}}(A_L-A_R), \qquad
   S_b = \frac{1}{\sqrt{2}}(B_L+B_R), \qquad  D_b = \frac{1}{\sqrt{2}}(B_L-B_R),
$$
together with the corresponding number operators

$$
  N_S= S_a^\dagger S_a, \qquad N_D = D_a^\dagger D_a, \qquad M_S= S_b^\dagger S_b, \qquad M_D = D_b^\dagger D_b,
$$
which count the number of bosons having vanishing (S) or non-vanishing (D) momentum in the two species. The momentum-mode Fock basis $\{\ket{N_S,\,N-N_S,\,M_S,\,M-M_S}\}$ can be chosen as a new basis against which it is possible to expand the generic state

$$
    \ket{\psi}=\sum_{n_S=0}^{N}\sum_{m_S=0}^{M} w_{n_S,m_S} \,\ket{n_S,m_S}_S \ket{N-n_S,M-m_S}_D,
$$
where we have set $ \ket{n_S,N-n_S,m_S,M-m_S} = \ket{n_S,m_S}_S \ket{N-n_S,M-m_S}_D$ in order to emphasize the difference between zero and non-zero momentum quantum numbers.
As a consequence, the density matrix relevant to the state is

$$
 \hat{\rho} = \ket{\psi}\bra{\psi}= \sum_{n_S=0}^{N}\sum_{m_S=0}^{M}
 \sum_{n_S^\prime=0}^{N}\sum_{m_S^\prime=0}^{M}
 w_{n_S,m_S} w^*_{n_S^\prime,m_S^\prime} \,\ket{n_S,m_S}_S \ket{N-n_S,M-m_S}_D\,\,
  \prescript{\null}{D}{\bra{N-n_S^\prime,M-m_S^\prime}} \, \prescript{\null}{S}{\bra{n_S^\prime,m_S^\prime}}.
$$

\begin{figure}[h]
    \centering
    \includegraphics[width=0.8\columnwidth]{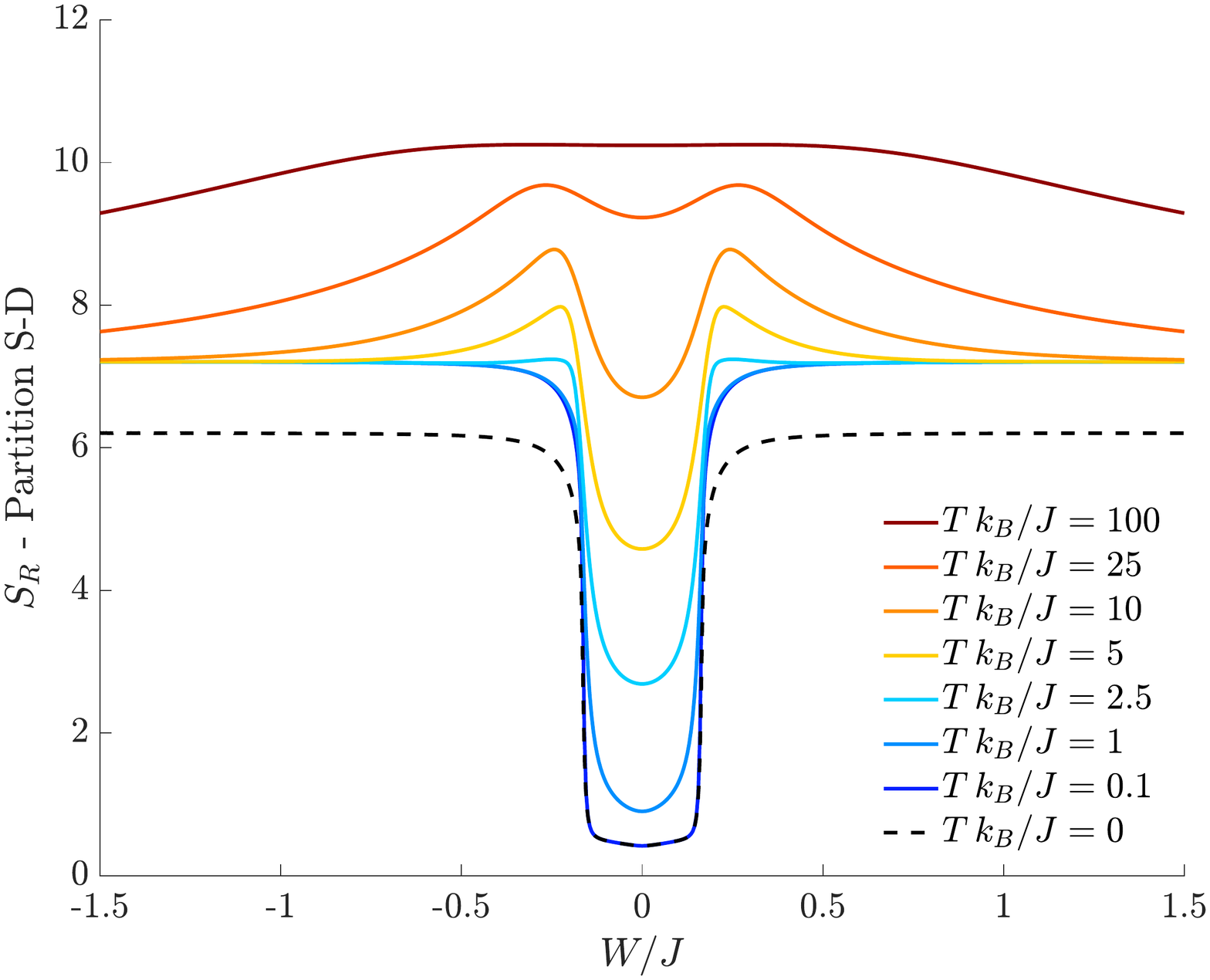}
    \caption{Bipartite residual entropy relevant to the partition S-D for different choices of the temperature. $N=30$, $M=40$, $J=1$, $U=0.1$, $k_B=1$.}
\label{fig:S_R_SD}
\end{figure}

The reduced density matrix relevant to the sub-system of bosons having non-vanishing momentum (modes D's) is obtained by tracing out the degrees of freedom relevant to the sub-system of bosons having zero momentum (modes S's)

$$
 \hat{\rho}_{D}=
 \sum_{n_S=0}^{N}\sum_{m_S=0}^{M}  \prescript{\null}{S}{\bra{n_S,m_S}} \hat{\rho}\ket{n_S,m_S}_S   =  \sum_{n_S=0}^{N}\sum_{m_S=0}^{M}
 |w_{n_S,m_S}|^2 \, \ket{N-n_S,M-m_S}_D \,\, \prescript{\null}{D}{\bra{N-n_S,M-m_S}}.
$$
For non-zero temperatures, one must consider the contributions of all the energy levels. Making use of the same scheme discussed in Subsection \ref{sec:Res_en_TSD}, as the reduced density matrix relevant to the thermal superposition is diagonal, the bipartite residual entropy (\ref{SR}) is found to be

$$
  S_R(\hat{\rho}_D)= - \sum_{n_S=0}^{N}\sum_{m_S=0}^{M}
 |C_{n_S,m_S}(T)|^2 \, \log_2 |C_{n_S,m_S}(T)|^2.
$$
Figure \ref{fig:S_R_SD} shows the bipartite residual entropy characterizing the separation between still and circulating bosons in respect of the ratio $W/J$, for different temperatures. At $T=0$, bipartite residual entropy corresponds to entanglement entropy and its plot (black dashed line) exhibits two sharp discontinuities at the two values of $W/J$ for which the DL phase transitions occur. Such discontinuities separate three quasi-plateaus corresponding to supermixed, mixed and demixed phases.

The central region (mixed species) features a quite small entanglement between circulating and still bosons. In fact, if the interspecies coupling $W$ is small compared to the tunneling $J$ and if the ratio $U/J$ is small enough to guarantee superfluid and delocalized bosons, momentum modes $S_a$ and $S_b$ are macroscopically occupied, while $D_a$ and $D_b$ are poorly populated. If the intraspecies repulsion $U$ tends to zero, one can show that the latter momentum modes are not populated at all, and, at $T=0$, the EE vanishes for $W/J=0$.

At finite temperatures, the behavior of the bipartite residual entropy still mirrors the presence of the three quantum phases. Unlike the behaviors of $S_R$ discussed in Subsection \ref{sec:Entanglement_LR}, where $S_R = 1$ associated to outer plateaus showed that system features two dominating space configurations, here, the value $S_R = 7.2$ implies that, for sufficiently large $|W|/J$, a much larger number of momentum configurations is involved in determining the system correlations.

Figure \ref{fig:S_R_SD} displays a gap between the plateau $S_R\approx 6.2$ obtained at $T=0$ (black dashed lines) and the limiting value $S_R\approx 7.2$ of the plateaus obtained at $T\neq 0$ (colored lines). This is due to the fact that, in the tails, the energy gap between the ground state and the first excited level becomes vanishingly small but remains non-zero and so the lowest non-zero temperature $T=0.1 J/k_B $ considered in Figure \ref{fig:S_R_SD} is already enough to populate both the ground state and the first excited level. The activation of the excited level (absent at $T=0$) is sufficient to redistribute the boson population thus causing the jump of $S_R$ from $6.2$ to $7.2$.  As noticed for the partition in terms of spatial modes discussed in Section \ref{sec:Entanglement_LR}, i) the maximum value of $S_R$ tends to the extreme value $\log_2 1271 \approx 10.31$ at high temperature and ii) given a certain value of $W/J$, the bipartite residual entropy steadily increases with temperature $T$ because more and more energy eigenstates become statistically accessible.

\subsection{Bipartite residual entropy for a partition characterized by boson species}
A third way to compute the bipartite residual entropy consists in partitioning the system in terms of species-A and species-B bosons. We use the representation in terms of space-mode Fock states, although the momentum-mode Fock basis is equally convenient to the job. Starting from density matrix (\ref{eq:Density_matrix_sites}), the reduced density matrix relevant to species-B sub-system is obtained by tracing out the degrees of freedom relevant to species-A sub-system

$$
  \hat{\rho}_{B} = \sum_{k=0}^{N}
  \, \prescript{\null}{L}{\bra{k}}\,\prescript{\null}{R}{\bra{N-k}}\hat{\rho} \ket{k}_L \ket{N-k}_R  =  \sum_{j=0}^{M}\sum_{j^\prime=0}^{M} \,\, C_{j,j^\prime}  \,\,\, \ket{j}_L\ket{M-j}_R  \,\,    \prescript{\null}{L}{\bra{j^\prime}}\,\,\prescript{\null}{R}{\bra{M-j^\prime}},
$$
where we have defined

$$
   C_{j,j^\prime} =  \sum_{n=1}^D \sum_{k=0}^{N} \, \frac{e^{-\beta E_n}}{\mathcal{Z}} w_{k,j,n} w_{k,j^\prime,n}^* .
$$
The diagonalization of $\hat{\rho}_{B}$ provides the eigenvalues $\{\lambda_j\}$ necessary to compute the relevant Von Neumann entropy

$$
  S_R(\hat{\rho}_B)=-\sum_{j=1}^{M+1} \lambda_j \,\log_2 \lambda_j.
$$

\begin{figure}[h]
    \centering
    \includegraphics[width=0.85\columnwidth]{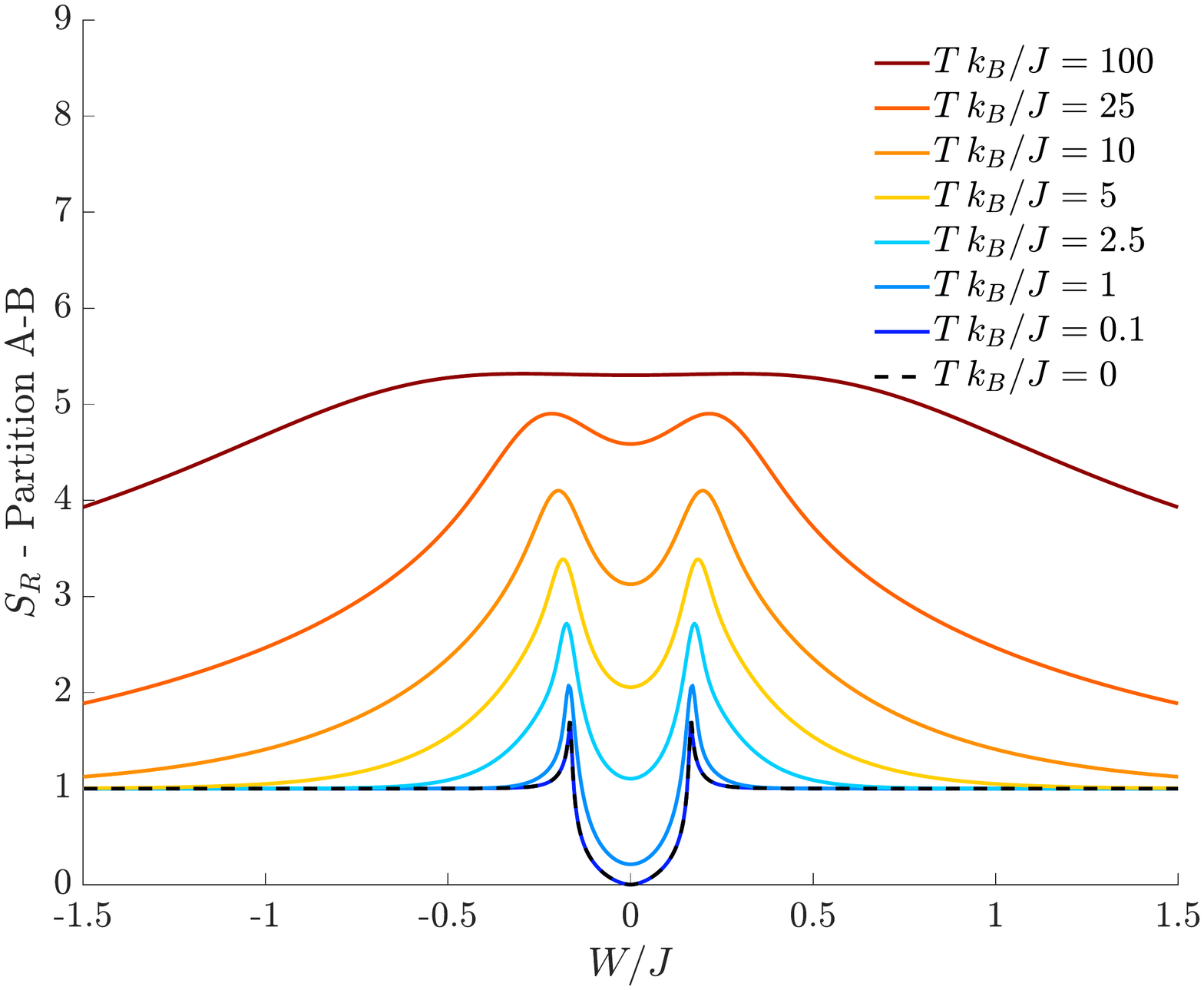}
    \caption{Bipartite residual entropy relevant to the A-B partition for different choices of the temperature. $N=30$, $M=40$, $J=1$, $U=0.1$, $k_B=1$.}
    \label{fig:S_R_AB}
\end{figure}

Figure \ref{fig:S_R_AB} shows the bipartite residual entropy relevant to species-mode partition scheme as a function of $W/J$, for different temperatures.

As in Figure \ref{fig:S_R_LR}, at zero temperature (black dashed line), two sharp peaks, at which the DL transitions occur, separate the three regions corresponding to the supermixed, mixed and demixed phase. Also in the present case, the outer regions consist of two quasi-plateaus whose height quickly converges to $1$, a limiting value which is, once again, reminiscent of the two-component character of cat states (\ref{eq:cats}) (recall that $1=2\times \left(-\frac{1}{2}\log_2 \frac{1}{2}\right)$).  As noted in the previous Subsections, one can show that the zero-temperature EE relevant to the space-mode and the momentum-mode separation schemes features a central minimum tending to zero for $J \to 0$ and $U \to 0$, respectively. In the current case, where the species-mode separation is adopted, the vanishing of the minimum of $S_R$ is obtained when the two species are non interacting, namely, for $W=0$.

When the temperature is switched on, the DL phase transitions become less abrupt and the corresponding peaks in the plots are less sharp. However, as shown in Figure \ref{fig:S_R_AB}, $S_R$ still represents an effective indicator of the critical behavior in a non-small temperature range. As for the $S_R$ analyzed in Section \ref{sec:Entanglement_LR}, the residual-entropy plot at non-zero temperatures shows that $S_R \to 1$ for $|W|/J$ large enough. Once more, the limiting value $S_R= 1=\log_2 2$ (which all colored lines of Figure \ref{fig:S_R_AB} converge to) highlights how the system features two equiprobable dominating configurations for large interactions. A non-vanishing $T$ disturbs the formation of such configurations since, in the tails, for a given value of $W/J$, the higher the temperature, the more $S_R$ differs from $S_R=1$. As for the other partition schemes, for large $T$, $S_R$ tends to a maximum value, $\log_2 D_B\simeq 5.36$, where $D_B=(M+1)$ is the dimension of sub-system-B Hilbert space.

\subsection{Bipartite residual entropy at zero temperature}
\label{sec:SR_0}
\begin{figure}[h]
    \centering
    \includegraphics[width=\textwidth]{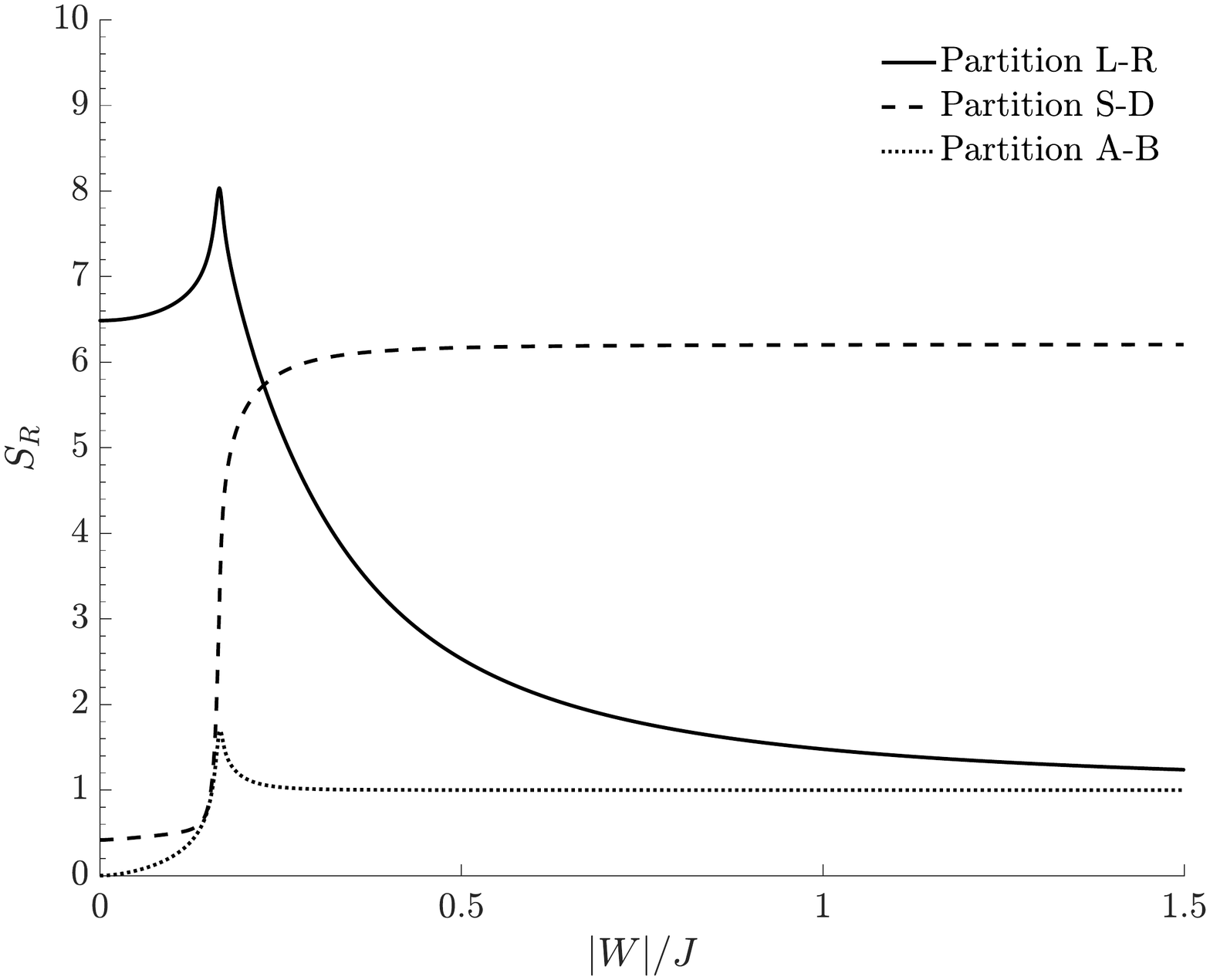}
    \caption{Residual entropies at $T=0$ as a function of $|W|/J$ for three different partitions of the Hilbert space: partition $L$-$R$ (continuous line), partition $S$-$D$ (dashed line), and partition $A$-$B$ (dotted line). $N=30$, $M=40$, $J=1$, $U=0.1$.}
    \label{SR0}
\end{figure}

As repeatedly stressed in the previous discussion, in principle, the choice of the partition, is completely arbitrary and independent on the system under examination. It has more to do with the concepts of ``observer'' and ``measure'' than with the physical system itself, opening interesting questions on the relation between entropy and quantum information.
To emphsize this fact, in Figure \ref{SR0} we compare the bipartite residual entropy at $T = 0$ for the three partition schemes considered above and shows how the presence of a non-zero bipartite residual entropy (i.e. of the EE in the ground-state) strongly depends on the choice of the partition. In particular, we notice how a strong entanglement in a partition can result in a weak (or zero) entanglement in another one. This is the case, e.g., of $W/J=0$ in which the ground-state is strongly entangled if measured through the partition $L$-$R$ (finite bipartite residual entropy $S_R$), or completely disentangled if measured through the partition $A$-$B$ (bipartite residual entropy $S_R=0$). In other words, in the same physical system, while the knowledge of the state of the system in the left (right) well is strongly correlated with the information on the state in the right (left) well, on the opposite, the knowledge of the species-A state does not produce information on the species-B state.

\section{Calculation of the EE in the coherent-state picture}
\label{sec:Coherent_States}
The coherent-state variational approach has found large application in the study of
many-body quantum systems \cite{ZFG} since, due to their semi-classical character, they  provide an effective description of physical systems and allow one to gain insights into their properties. Also, from the experimental point of view, coherent states have an important role since their semi-classical character enables one to achieve a realistic approximation of the quantum state describing the real system.

An su(2) coherent state describing single condensate trapped in a dimer is given by \cite{JPA41}

\begin{equation}
    \ket{\xi_L, \xi_R} =\frac{1}{\sqrt{N!}}\left(\xi_L A_L^\dagger + \xi_R A_R^\dagger  \right)^N \ket{0},
\label{CS1}
\end{equation}
where $\ket{0} = \ket{0, 0}$ is the boson vacuum state  and the normalization condition $|\xi_L|^2+|\xi_R|^2=1$ must be assumed. Since $\bra{\psi_a} A_\sigma^\dagger A_\sigma \ket{\psi_a} = N |\xi_\sigma|^2$, with $\sigma =R, L$, is the expectation value of number operator $N_\sigma = A_\sigma^\dagger A_\sigma$ then $|\xi_\sigma|^2$ represents the fraction of bosons in the well $\sigma$.  In the following, we employ combinations of coherent states (\ref{CS1}) (for a single species in a double well) to approximate the cat structure of the ground state relevant to the TSD system in the strong-interaction regime, both for $W/J >0$ and for $W/J <0$.

\begin{enumerate}
\item Supermixing (attractive cat). If the interspecies attraction ($W/J <0$) is large enough, the two species aggregate together in the same well. Since none of the two wells is privileged with respect to the other, quantum mechanically both configurations are equally probable, and the system lives in both states at the same time. By using the notation of Formula (\ref{CS1}), the resulting cat state can be written as

$$
   \ket{\Psi}= \frac{1}{\sqrt{2}} \biggl[\ket{\text{Loc}}_{a,L}  \ket{\text{Loc}}_{b,L} + \ket{\text{Loc}}_{a,R} \ket{\text{Loc}}_{b,R} \biggr]
   =
 \frac{1}{\sqrt{2}} \biggl[\ket{\lambda_a,\eta_a} \ket{\lambda_b,\eta_b} +\ket{\eta_a,\lambda_a} \ket{\eta_b,\lambda_b} \biggr],
$$
where "Loc" stands for "localized" and entails the fact that $|\eta_c|^2 \ll |\lambda_c|^2$.
Following the scheme discussed in Ref. \cite{PRA87}, one can show that the expectation value of the model Hamiltonian reduces to

$$
 E(\lambda_a,\eta_a,\lambda_b,\eta_b) = \frac{U}{2}N(N-1)\left(|\lambda_a|^4 + |\eta_a|^4 \right) -2JN (\mathrm{Re}\{\lambda_a\eta_a\})
$$
$$
+ \frac{U}{2}M(M-1)\left(|\lambda_b|^4 + |\eta_b|^4 \right) -2JM (\mathrm{Re}\{\lambda_b\eta_b\})
+ W\left(|\lambda_a|^2|\lambda_b|^2+|\eta_a|^2|\eta_b|^2 \right)
$$
where the local order parameters $\lambda_a$, $\lambda_b$, $\eta_a$, and $\eta_b$ are complex quantities defined as

$$
  \lambda_a = \sqrt{1-x_a}e^{i \theta_a},\qquad \eta_a = \sqrt{x_a}e^{i\phi_a}, \qquad    \lambda_b = \sqrt{1-x_b}e^{i\theta_b},\qquad \eta_b = \sqrt{x_b}e^{i\phi_b}.
$$
The minimum-energy configuration is reached for $\phi_a=\theta_a$, $\theta_b= \phi_b$ and

$$
   x_a = \frac{J^2}{(N U-U+M W)^2}, \qquad x_b = \frac{J^2}{(MU-U+ N W)^2}.
$$
These formulas give the fraction of bosons characterizing the minority component and, correctly, give zero in the limit $W\to-\infty$.

\medskip

\item{Demixing (repulsive cat).} If the interspecies repulsion ($W/J >0$) is large enough, the two condensed species separate in different wells. Similarly to what explained in the previous paragraph, the ground state features a two-sided cat-like structure, because left (right) well can indistinctly host species A (B). Hence, the quantum state consists of an equally-weighted superposition of the two possible arrangements

$$
   \ket{\Psi}= \frac{1}{\sqrt{2}} \biggl[\ket{\text{Loc}}_{a,L}  \ket{\text{Loc}}_{b,R} + \ket{\text{Loc}}_{a,R} \ket{\text{Loc}}_{b,L} \biggr] = \frac{1}{\sqrt{2}} \biggl[ \ket{\lambda_a,\eta_a}\ket{\eta_b,\lambda_b}+ \ket{\eta_a,\lambda_a} \ket{\lambda_b,\eta_b} \biggr],
$$
where $\lambda_c$, $\eta_c$ are such that $|\eta_c|^2\ll |\lambda_c|^2$ and (obviously) $|\lambda_c|^2+|\eta_c|^2=1$, with $c=a,b$.
Following the variational approach described in the previous paragraph, and adopting the same conventions, we obtain that the variational energy is minimized for $\theta_a=\phi_a$, $\theta_b= \phi_b$ and

$$
   x_a = \frac{J^2}{(NU-U-M W)^2}, \qquad x_b = \frac{J^2}{(M U-U-NW)^2}
$$
Parameters $x_a$ and $x_b$ represent the fractions of bosons which do not aggregate with the others and thus make the "demixed phase" not ideal. Notice that, correctly, if $W\to +\infty$, then $x_{a,b}\to 0$, i.e. the demixing gets more and more complete.

Both for the supermixing and for the demixing scenario, after computing the fraction of bosons in each well, it is possible to reconstruct the cat state by superimposing two coherent states. This procedure, described in Appendix \ref{app:S_R_coherent_states}, allows one to analytically compute the EE between left-well and right-well bosons, at zero temperature \cite{PRA87}. As shown in Figure  \ref{fig:Confronto_stati_coerenti}, the result perfectly matches the numerical EE, of course in the validity range of this approximation, i.e. in the whole range of $|W|/J$ except the central region (mixed phase) between the two critical values.

\begin{figure}[h]
    \centering
    \includegraphics[width=0.85\columnwidth]{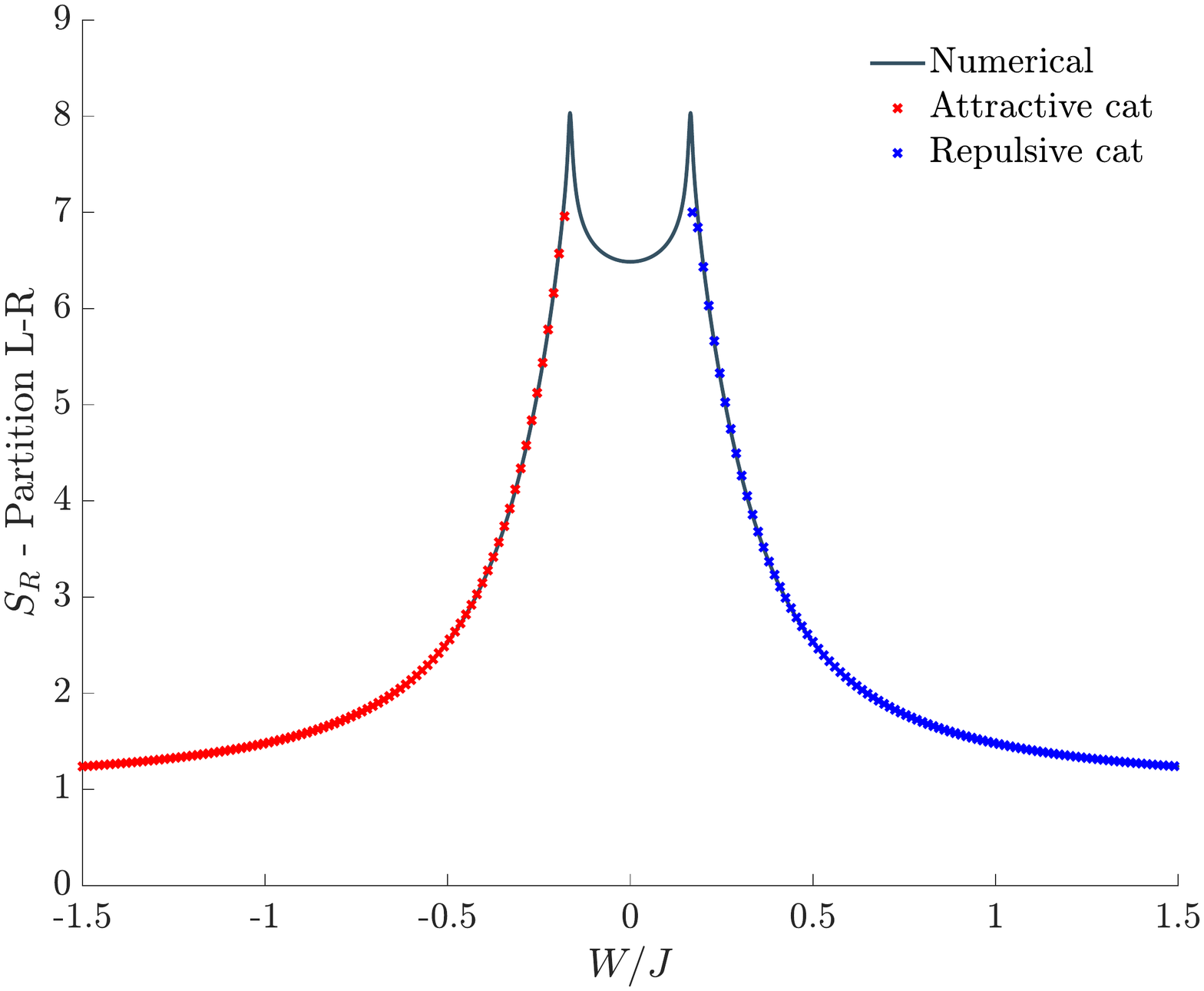}
    \caption{Entanglement entropy between left-well bosons and right-well bosons: comparison between the results derived within the coherent-state approach and the numerical ones.}
    \label{fig:Confronto_stati_coerenti}
\end{figure}

\end{enumerate}

\section{Calculation of the bipartite residual entropy in a restricted energy basis}
\label{sec:Restricted}
As already explained, the density operator associated to a thermal mixture of eigenstates is

$$
	\hat{\rho}=\frac{1}{\mathcal{Z}} \sum_{n=1}^D e^{-\beta E_n} \ket{\psi_n}\bra{\psi_n}
$$
where $E_n$ is the energy eigenvalue associated to the energy eigenstate $\ket{\psi_n}$, $\beta$ is (proportional to) the inverse temperature and $D$ is the dimension of the Hilbert space of physical states. From a computational, but also from a conceptual point of view, $\hat{\rho}$ is the superposition of $D$ different contributions, each one weighted by a different Boltzmann factor. The dimension $D$ rapidly increases with the number of particles hosted in the system its exact value being $D=(N+1)(M+1)$. As a consequence, the computation of the thermal density matrix becomes unfeasible even for a relative small number of bosons. By taking advantage of the well-known equivalence between microcanonical and canonical ensemble (see, e.g. \cite{KersonHuang}), for large numbers of particles, we provide an effective way to approximate a thermal state. To this end, we consider just a restricted set of energy eigenstates, namely those $\ket{\psi_n}$ whose energy $E_n$ lies in the range $[\avg{E}-\sigma_E; \avg{E}+\sigma_E]$ where

$$
   \avg{E} = \frac{1}{\mathcal{Z}} \sum_{n=1}^D E_n e^{-\beta E_n}, \qquad  \sigma_E = \sqrt{\avg{E^2}-\avg{E}^2}
$$
are the expectation value of the energy and its standard deviation, respectively.
The density matrix relevant to this restricted thermal state is thus constructed by equally-weighting the contributions coming from such $\ket{\psi_n}$, i.e.

$$
   \hat{\rho}_{restricted} = \frac{1}{N_*} \sum_n^* \ket{\psi_n}\bra{\psi_n}
$$
where $N_*$ is the number of energy eigenstates whose energies $E_n$ lie in the aforementioned interval.

\begin{figure}[h]
    \centering
    \includegraphics[width=0.85\columnwidth]{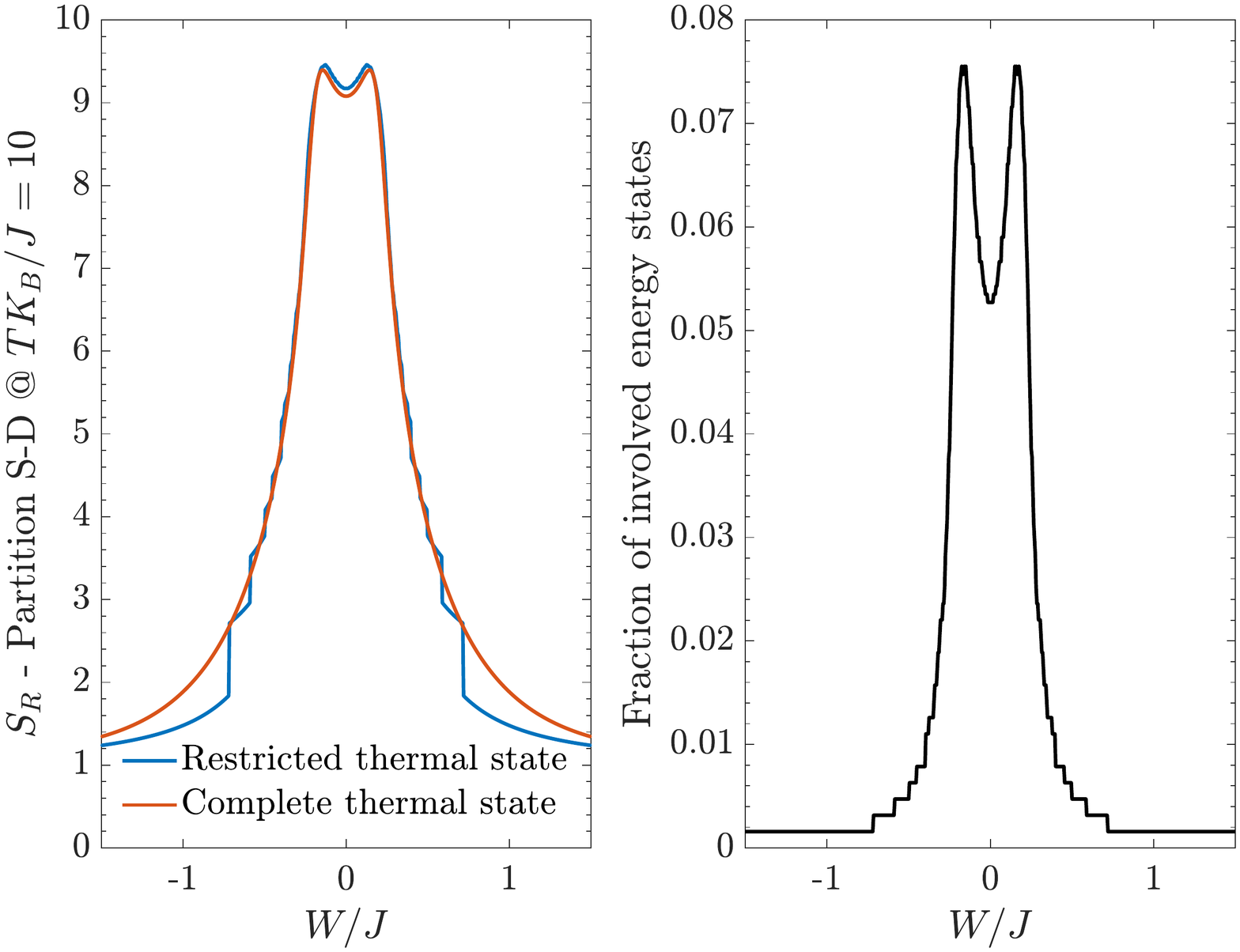}
    \caption{Bipartite residual entropy relevant to the partition L-R. Left panel: comparison between the results obtained within the restricted set of eigenstates and the exact ones. Right panel: fraction of energy eigenstates which takes part in the restricted thermal density matrix.}
    \label{fig:Comparison_restricted_LR}
\end{figure}

To test the effectiveness of the bipartite residual entropy as a critical indicator, we consider the partition in terms of left-well bosons and right-well bosons, we set a non-zero value of the temperature and we compare the results obtained from a complete and from a restricted thermal state. The left panel of Figure \ref{fig:Comparison_restricted_LR} shows an overall good agreement between such results, especially in the central region (small $|W|/J$ values), while the outermost regions feature step-like discontinuities. The presence of such discontinuities can be understood observing ethe right panel of Figure \ref{fig:Comparison_restricted_LR}, which shows the fraction of energy states involved in the restricted thermal state, $N_*/D$, as a function of $W/J$. As $W/J$ increases, in fact, fewer and fewer energy states join the restricted thermal state and their inherently discrete character is reflected by the presence of step-like regions, each one corresponding to the activation of a single energy state.

\section{Conclusions}
\label{sec:Conclusions}
In this work, we have investigated the equilibrium and the bipartite residual entropy in a two-species Bose-Hubbard dimer at zero and non-zero temperature. In Section \ref{sec:the_model} we have introduced the model and highlighted the importance of $W$ (the interspecies repulsion) in determining the quantum phase of the system (supermixed for $W/J\ll 0$, mixed for small $|W|/J$ and demixed for $W/J\gg 0$). In Section \ref{sec:S_eq_S_R} we have introduced the concepts of equilibrium and bipartite residual entropy commenting on the fact that, at zero temperature, the latter corresponds to the entanglement entropy.

Section \ref{sec:SR_T} has been devoted to the analysis of the bipartite residual entropy for three different partitions of the total system. In this regard, we have stressed the fact that different ways of partitioning the system into two sub-systems, correspond to different kinds of residual entropies $S_R$. In all three cases, $S_R$ features discontinuities where the localization-delocalization phase transitions occur and quasi-plateaus where two dominating macroscopic configurations emerge. Bipartite residual entropy is therefore a valid critical indicator not only at zero temperature (where it corresponds to the entanglement entropy, a purely quantum correlation), but also at higher temperatures, where it is influenced by the classical correlation between the sub-systems and by classical entropy. Interestingly, we have evidenced that, at zero temperature, \textit{i)} a non-zero hopping $J$  causes a non-zero entanglement between spatial modes, \textit{ii)} the intraspecies interaction $U$ contributes to the entanglement between momentum modes, and \textit{iii)} the interspecies interaction $W$ is responsible for the entanglement between species modes.

In Section \ref{sec:Coherent_States}, we have introduced su(2) coherent states and developed a fully-analytic variational approach apt to describe the supermixed and the demixed phases at zero temperature. The superposition of two such coherent states has provided a good approximation of the ground state of the system in a non-small range of $W/J$, as demonstrated by the comparison with the numerical results. In Section \ref{sec:Restricted} we have approximated the complete thermal superposition (\ref{eq:thermal_superposition}) with an incoherent combination of a reduced number of equally-weighted energy eigenstates and showed that the bipartite residual entropy is still a good critical indicator, well reproducing the exact results obtained numerically.

\appendix
\section{Entanglement entropy and coherent states}
\label{app:S_R_coherent_states}
On the basis of the coherent-state approach derived in Section \ref{sec:Coherent_States} and in the same spirit of Ref. \cite{PRA87}, we compute the entanglement entropy between left-well bosons and right-well bosons. To begin, let us define $\rho_{n,m} (i)$ as the probability of having $n$ bosons of species A and $m$ bosons of species B at site $i$. The normalization of probability requires that

$$
\sum_{n=0}^{N}\sum_{m=0}^M \,\rho_{n,m}(i) =1
$$
where $N$ is the total number of bosons of species A and $M$ is the total number of bosons of species B. Let us define the single site entropy $S_i$ as follows:

$$
  S_i = -\sum_{n=0}^N\sum_{m=0}^M \,\rho_{n,m} (i) \, \log_2 \rho_{n,m} (i)
$$
Neglecting the possible presence of cat states (a situation that will be re-inserted a posteriori), a generic coherent state can be written in the factorized form

$$
  \ket{\Psi}=\left[ \frac{1}{\sqrt{N!}}\left( \xi_L A_L^\dagger +\xi_R A_R^\dagger \right)^N \ket{0} \right ] \left[ \frac{1}{\sqrt{M!}}\left( \nu_L B_L^\dagger +\nu_R B_R^\dagger \right)^M \ket{0} \right ]
$$
Of course the normalization conditions $   |\xi_L|^2+|\xi_R|^2=1, \quad |\nu_L|^2+|\nu_R|^2=1 $ must hold. State $\ket{\Psi}$ can be recast into the form

$$
   \ket{\Psi}=\left[\sum_{n=0}^N \frac{\sqrt{N!}}{n!\,(N-n)!} \xi_L^n \left(A_L^\dagger \right)^{n} \xi_R^{N-n}\left(A_R^\dagger\right)^{N-n} \, \ket{0} \right]   \left[\sum_{m=0}^M \frac{\sqrt{M!}}{m!\,(M-m)!} \nu_L^m \left(B_L^\dagger \right)^{m} \nu_R^{M-m}\left(B_R^\dagger\right)^{M-m} \, \ket{0} \right]=
$$
$$
   = \left[\sum_{n=0}^N \frac{\sqrt{N!}}{\sqrt{n!}\,\sqrt{(N-n)!}} \xi_L^n \xi_R^{N-n} \, \ket{n,N-n}_a \right]   \left[\sum_{m=0}^M \frac{\sqrt{M!}}{\sqrt{m!}\,\sqrt{(M-m)!}} \nu_L^m  \nu_R^{M-m}\, \ket{m,M-m}_b \right]
$$
We calculate the reduced density matrix $\rho$ partitioning the system into two sub-systems (left-well bosons and right-well bosons) and tracing out the degrees of freedom relevant to one of them. For example

$$
    \rho = \sum_{n=0}^{N}\sum_{m=0}^M  \null_R\braket{n,m}{\Psi} \braket{\Psi}{n,m}_R
$$
Taking into account the orthogonality of the states, the reduced density matrix which originates from a coherent state can be written as

$$
    \rho_{n,m}= \frac{N!M!}{n!m!(N-n)!(M-m)!} \xi_L^n \xi_R^{(N-n)}\nu_L^m \nu_R^{(M-m)}(\xi_L^*)^n (\xi_R^*)^{(N-n)}(\nu_L^*)^m (\nu_R^*)^{(M-m)}=
$$
$$
   = \left[\binom{N}{n} |\xi_L|^{2n}(1-|\xi_L|^{2})^{(N-n)} \right]\left[ \binom{M}{m} |\nu_L|^{2m}(1-|\nu_L|^{2})^{(M-m)}\right]
$$
where the expressions of coefficients $\xi_L=\xi_L(J,U,W)$, $\xi_R=\xi_R(J,U,W)$, $\nu_L=\nu_L(T,U,W)$ and $\nu_R=\nu_R(T,U,W)$ can be computed within the variational approach.
In passing, notice that the probability distribution is correctly normalized, i.e. $\sum_{n=0}^N\sum_{m=0}^M \rho_{n,m}=1$.
The Von Neumann entropy of the remaining sub-system can be thus computed as

$$
  S = -\sum_{n=0}^N\sum_{m=0}^M \,\rho_{n,m}  \, \log_2 \rho_{n,m}
$$

This quite general procedure needs to be slightly modified in case one is considering cat states. In fact, the reduced density matrix must take into account the two-sided nature of a cat state and so it must be written as the average of the densities matrices relevant to simple coherent states, namely

$$
  \rho_{\text{side\,L}_{n,m}}=  \left[\binom{N}{n} |\xi_L|^{2n}(1-|\xi_L|^{2})^{(N-n)} \right]\left[ \binom{M}{m} |\nu_L|^{2m}(1-|\nu_L|^{2})^{(M-m)}\right]
$$
$$
  \rho_{\text{side\,R}_{n,m}}=  \left[\binom{N}{n} |\xi_L|^{2(N-n)}(1-|\xi_L|^{2})^{n} \right]\left[ \binom{M}{m} |\nu_L|^{2(M-m)}(1-|\nu_L|^{2})^{m}\right]
$$
implying
$$
  \rho_{\text{cat}_{n,m}}= \frac{1}{2}\left[ \rho_{\text{side\,L}_{n,m}}  +  \rho_{\text{side\,R}_{n,m}}\right]
$$

\section{Quasi-degeneracy of the ground-state}
\label{appB}
\begin{figure}[h]
    \centering
    \includegraphics[width=0.7\textwidth]{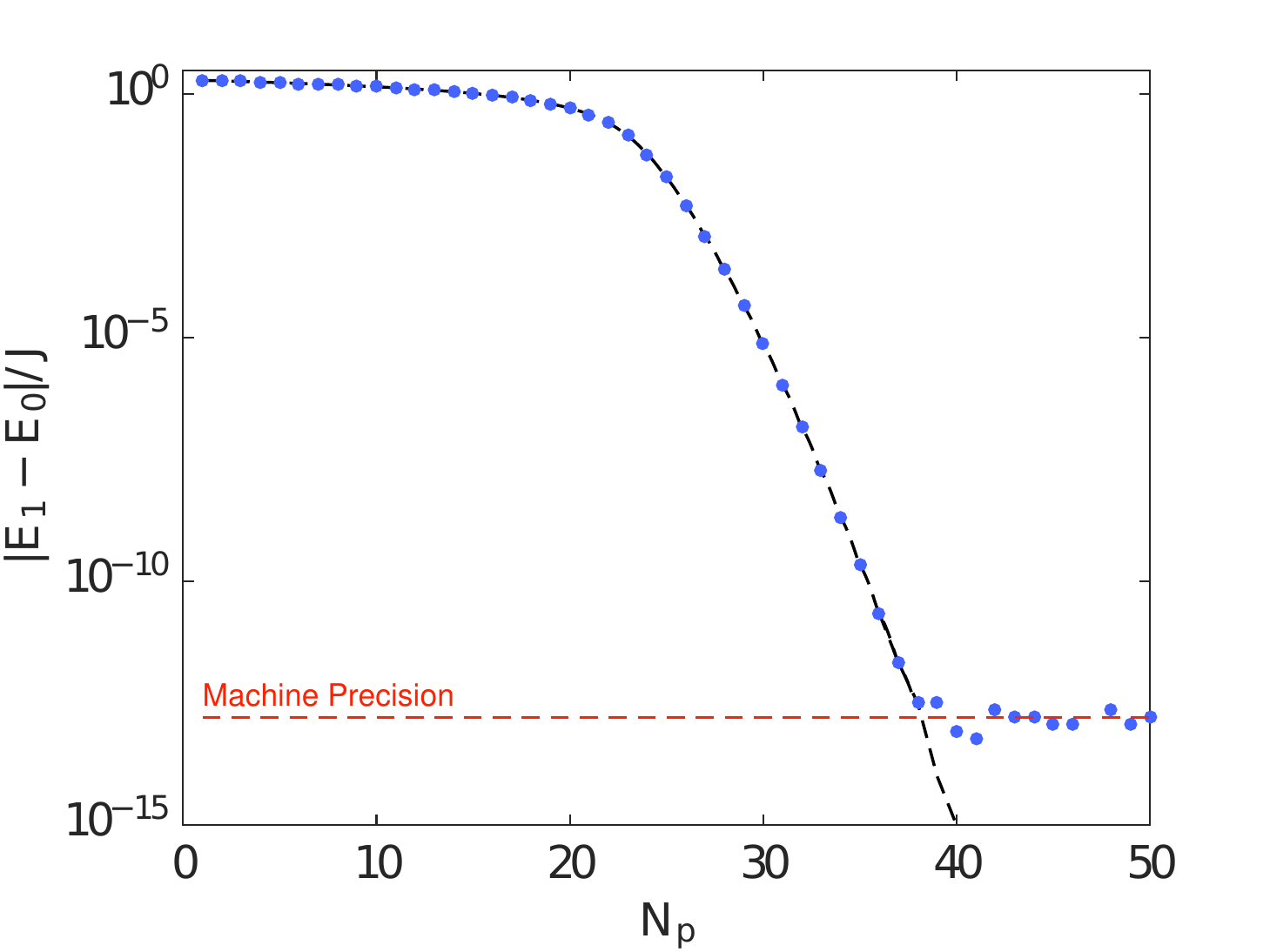}
    \caption{Splitting between the ground-state energy $E_0$ and the first excited state energy $E_1$ as function of the number of particle per species ($N=M=N_p$), $J=1$, $U=0.1$, $W=0.2$.}
    \label{deE01}
\end{figure}

Due to the spectral-collapse \cite{JPB49,PRE95}, for sufficiently strong values of $W/J$, the TSD ground-state may appear quasi-degenerate. However the degeneracy of the ground-state is only apparent as Hamiltonian (\ref{model}) is non-degenerate \cite{Landau, Wei2014}. As shown in Figure \ref{deE01} the energy splitting between the ground-state energy $E_0$ and the first energy level $E_1$ decays exponentially as function of the number of particles per species ($N$ and $M$).
Energy levels $E_0$ and $E_1$ differs always by a small, finite, quantity function of the interactions and the number of particles. This has been verified in Figure \ref{deE01} down to computational limit fixed by the machine precision.




\end{document}